\documentstyle[aps,prl,multicol,epsfig]{revtex}
\begin{document}
\title{Dynamical properties of the pinned Wigner crystal}
\author{R. Chitra{$^1$}, T. Giamarchi{$^1$} and P. Le Doussal{$^2$}}
\address{{$^1$} Laboratoire de Physique des Solides, UPS Bat. 510, 91405
Orsay, France. \cite{cnrs}}
\address{{$^2$}CNRS-Laboratoire de Physique Theorique de
l'Ecole Normale Superieure,
24 rue Lhomond,75231 Cedex 05, Paris, France.
}

\maketitle

\begin{abstract} {
We study various dynamical properties of the  weakly pinned Wigner
crystal in a high magnetic field.
Using a Gaussian variational method we can compute the full
frequency and field dependence of the
real and imaginary parts of the diagonal and Hall conductivities.
The zero temperature Hall resistivity is independent of frequency and
remains unaffected by disorder at its classical value.
We show that, depending on the inherent length
scales of the system, the pinning peak and the threshold electric
field exhibit strikingly different magnetic field dependences.}

\end{abstract}

\begin{multicols}{2}

Though the exotic possibility of electron crystallisation was discussed
decades ago by Wigner \cite{EW}, its experimental realization has been a
challenge due to difficulties in obtaining sufficiently low
density electron systems. However, this problem can  be circumvented by
subjecting the $2$DEG
to large   magnetic fields  which  facilitate
crystallisation of  even dense electron systems.
The quest for the Wigner crystal (WC) in   mono \cite{WC1,WS,NC1,WC2} and
bilayer \cite{BWC} quantum Hall samples
indicated the existence of  a quaint insulating state at
filling fractions where crystallisation was  theoretically expected.
These and other detailed studies \cite{SL}
of the insulating state revealed
that the diagonal resistivity $\rho_{xx}$  diverges as  the temperature
$T\to 0$
and shows activated behaviour at
finite $T$
whereas the Hall resistivity $\rho_{xy}$  is temperature independent and
has a value
close to the classical Hall value.

Although measurements of activated linear and non-linear dc conductivity
and the luminescence spectrum of radiative
recombination \cite{KT} were consistent with interpretations in terms
of a pinned WC, the finite value seen for $\rho_{xy}$ was
unexpected. This prompted other interpretations of the observed
insulating phase \cite{Klit,KLZ} and in particular the existence of
a new phase, the Hall insulator (HI)
\cite{KLZ}. The Hall insulator is defined as a phase where
$\lim_{\omega \to 0} {{\sigma_{xy}(\omega)} \over {\omega^2}} =~\rm{cste}$
which yields $\rho_{xx} \to \infty$ and a finite
$\rho_{xy}$ in the limit $T\to 0, \omega \to 0$. This
was proved only for {\it non interacting electrons} in a random potential
(i.e an Anderson insulator in presence of a magnetic field)
and qualitative arguments suggested that it holds
for interacting systems as well. However none of these arguments
take into account the possible local crystalline order
which could result in radically different physics
as compared to the disordered electron fluid. Indeed
periodicity  plays an important
role in other disordered systems, such as vortex lattices
\cite{TGPL}.

It is thus of prime importance to investigate in detail the
transport properties of a pinned WC.
Transport properties are especially important
here because of the extreme difficulty of a
direct experimental verification of local
crystalline order.
One of  the few theoretical attempts made to
predict these properties  was that of Ref.\onlinecite{FL}
where the  related  problem of charge density
waves (CDW) in a magnetic field
was studied.
The harmonic approximation used, however, did not allow the extraction
of the detailed frequency dependence of
the conductivities. Later works \cite{NLM} focussed
on the sliding state and the effects of
free carriers, or on the effect of strong disorder
\cite{WKS}.
A noteworthy point is that none of these calculations
consider both a lattice structure and
modulation of disorder at scales smaller than the lattice
spacing
\cite{TGPL}.
This feature which is  absent in CDW
turns out to play a crucial role in the physics of the WC.

In this Letter, we compute
for the first time the real and imaginary parts of the frequency dependent
conductivities of a weakly pinned WC.
We find that even some of the
main features derived in Ref.\onlinecite{FL}
are incorrect. The results we
obtain provide a basis for comparison with recent experiments
which map out the low frequency behaviour of the
conductivity\cite{SL1,CC}.

Our starting point is the WC in a magnetic field $B$
 with lattice spacing $a$ modelled by
 an elastic hamiltonian
\cite{FL}.
The electrons   at site $i$ are displaced  from their
mean  equilibrium positions ${\bf R}_i$
by ${\bf u}({\bf R}_i,t)$.
We also take into account   the Coulomb repulsion  between
density fluctuations.
We use  the following
 decomposition
${\bf u}({\bf q}) = \hat {\bf q} u_L({\bf q}) + {\hat {\bf q}}
\wedge {\hat{\bf z}} u_T({\bf q})$
where $L,T$ denote the longitudinal and transverse components.
The corresponding action  in the imaginary time formalism
is
\begin{eqnarray} \label{ham}
S[u]& =& \int_ {\bf q}  \sum_n  [ u^L_{q,\omega_n}(\rho_m\omega_n^2 +cq^2
+dq)u^L _{-q,
-\omega_n}\nonumber\\
&+&u^T_{q,\omega_n} (\rho_m \omega_n^2 + c q^2 )u^T_{-q,
-\omega_n}
 + \rho_m\omega_c \omega_n (u^L_{q,\omega_n} u^T_{-q ,-\omega_n}
\nonumber \\
&- &
u^L_{-q ,-\omega_n} u^T_{q ,\omega_n)}]
+  \int_{\bf x} \int_0^{\beta \hbar} d\tau W(x) \rho(x,\tau)
\end{eqnarray}
\noindent
$\rho_m,\rho_c$ are the mass and charge densities.
$c$ and $d$ are the shear and bulk modulus respecitvely.
For the WC, the  presence of
 Coulomb forces
 results \cite{NLM,BM} in a bulk modulus   $d=
{\rho_c^2 \over \epsilon_0}$ much greater than the shear modulus
 $ c= {{\rho_c^2 a} \over \epsilon_0}$ ($\epsilon_0$ is the dielectric
constant of the substrate).
$\omega_c={{\rho_c B} / {\rho_m}}$
is the cyclotron frequency and
the Matsubara frequencies at temperature $T$ are $\omega_n={{2\pi n} /
{\beta \hbar}}$  where
$\beta={1 / T}$.
 From (\ref{ham}), we see that the Coulomb interaction affects only the
longitudinal
modes.
The magnetic field
couples the transverse and longitudinal modes.
$\langle \qquad \rangle$ denote  averages over quantum and thermal
fluctuations and
$\overline{\qquad}$  are disorder averages.
For the pure system the
quantum fluctuations  result in $\langle {\bf u}^2 \rangle \sim l_c^2$
 where $l_c$ is the
magnetic length $l_c= \sqrt {{\hbar} / {eB}}$.
The time-independent disorder potential $ W(\bf x)$  is short range
correlated (of range $r_f$)
$\overline{ W({\bf x}) W({\bf x}^\prime )} = \Delta \delta_{r_f} ({\bf
x}-{\bf x}^{\prime})$
and couples to the  density of electrons
 $\rho({\bf x},t)= \sum_i \delta({\bf x}- {\bf R}_i -{\bf u}({\bf
R}_i,t))$.
Using the decomposition of
the density into lattice harmonics
\cite{TGPL} (valid in the absence of topological defects)
and replicas to average over disorder
we obtain the effective action
\begin{eqnarray} \label{decadix}
S_{\text{eff}} =&& \sum_a H[u^a]
 - {1\over {2\hbar}} \int _{\bf  x} \int \int d\tau d\tau^{\prime}\nonumber
\\
&&\sum_{a,b,{\bf K}} \Delta_K \cos({\bf K}.({\bf u}^a ({\bf x}\tau) -{\bf
u}^b ({\bf x},\tau^{\prime})))
\end{eqnarray}
\noindent
 $a,b$ denote the replica indices, ${\bf K}$ are the reciprocal lattice
vectors  and $\Delta_K \sim \Delta \exp{-{\bf K}^2
 r_f^2}$. Note that it is important to retain all harmonics.
The  disorder averaging also yields a term quadratic in the displacements
which has been
absorbed by a shift in $\bf u$. This shift does not affect
the conductivity and we neglect it henceforth.
Actions similar to (\ref{decadix}) can be used to describe 3-d
classical problems such as vortex lattices
with correlated disorder \cite{TP} and long range interactions.

To study  this model we use  the
gaussian variational method (GVM).
This  quantitative method
 allows us  to  compute the Green functions and
  hence the conductivity of the
system described by (\ref{decadix}).
 Unlike  previously used methods
\cite{FL} the GVM  is self-consistent, has no undetermined adjustable
parameters
and  also incorporates important physical feautures of the
problem such as the existence of many metastable states.
It allows one to go beyond
simple  static arguments as will
be shown later.
We  introduce the variational action \cite{TP}
\begin{equation} \label{gham}
S_0= {1 \over 2} \int_{\bf q}  \sum_n u_{\alpha, q,\omega_n}^a
G^{{ab}-1}_{\alpha
\beta}(q,\omega_n) u_{\beta,-q, -\omega_n}^b
\end{equation}
\noindent
where the  Green functions
$G^{ab}_{\alpha \beta}$ are the variational parameters
( $\alpha,\beta=L,T$
and summations over repeated indices are implicit).
They are determined by  solving the self-consistent
saddle point equations obtained by extremizing the
variational free energy
$F_{\text{var}}=F_0 + \langle S_{\text{eff}}-S_0\rangle_0$.
The method extends the one used in \cite{TP}
and all technical details will be presented in
\cite{future}. As in \cite{TP} the solution
has a replica symmetry broken  structure
necessary to  correctly describe  the localization.
The final result \cite{future} is a closed set of
equations for the connected part of the Green function
$G^{-1}_{c \alpha \beta}= \sum_b G^{-1}_{\alpha \beta,ab}$
which determine all physical quantities of interest here.
These equations are respectively:
\begin{eqnarray} \label{gctl}
G_{cT}^{-1}&= &(cq^2 + {\rho_m}\omega_n^2) +F + {{{\rho_m^2}\omega_n^2
\omega_c^2} \over
{(cq^2+dq+ {\rho_m}\omega_n^2 +F)}} \\
G_{cL}^{-1}&= &(cq^2 + dq+{\rho_m}\omega_n^2) +F + {{{\rho_m^2}\omega_n^2
\omega_c^2} \over
{(cq^2 +\rho_m\omega_n^2 +F)}}
\nonumber \\
G_{cLT}^{-1}&= &{\rho_m\omega_n \omega_c} +{{(cq^2 +\rho_m\omega_n^2+F)
(cq^2+dq +\rho_m\omega_n^2+F)} \over {\rho_m\omega_n \omega_c}}  \nonumber
\end{eqnarray}
\noindent
with
$F= I(\omega_n) + \Sigma (1 - \delta_{n,0})$. The localized
phase is characterized by a non zero $\Sigma$, from which a
length scale $l$ can be defined
through $ \Sigma=   c   l^{-2}$. The function
$I(\omega_n)$ is defined as:
\begin{equation} \label{iwn}
I(\omega_n)= {2 \over \hbar}\int_0^{\beta} d\tau (1-\cos(\omega_n \tau))
(V^{\prime}(\tilde B(\tau)) - V^{\prime}(B))
\end{equation}
\noindent
where the local diagonal correlation $\tilde B(\tau) =
1/2 \overline{\langle (u(0,\tau)-u(0,0))^2\rangle }
=1/2 ({\tilde B}_L(\tau)+{\tilde B}_T(\tau))$
and the off diagonal part $B$ are given by:
\noindent
\begin{eqnarray} \label{btbl}
 {\tilde B}_{T,L}(\tau)&& = {{2\hbar } \over \beta} \int_q \sum_n G_{cT,L}
(1- \cos
(\omega_n \tau)) \nonumber \\
 B && = {\hbar \over \beta}\int_{\bf q} [\sum_{n\neq0} (G_{cL}(q,\omega_n)+
G_{cT}(q,\omega_n) )]+\nonumber  \\
&& {1 \over {cq^2 +dq+\Sigma}}+ {1 \over {cq^2 +\Sigma}}
\end{eqnarray}
\noindent
Finally the set of equations  close as $\Sigma$ is itself determined
by
\begin{eqnarray} \label{self1}
1&=&-2 V^{\prime \prime}(B)\int_{\bf q} {1 \over {( cq^2 + \Sigma)^2}} +
 {1 \over { (cq^2 + dq + \Sigma)^2}}
\end{eqnarray}
\noindent
The primes denote derivatives.
All information on the disorder is
contained in the auxiliary function
$V[B]= (4\hbar)^{-1} \sum_K \Delta_K  \exp(- K^2 B)$.

In this paper, we focus on the transport properties
but  other quantities such as positional
correlation functions can also be computed \cite{future,TP}.
The dynamical conductivities are given by
the standard analytical continuation of the
Green's functions $\sigma_{\alpha \beta}(\omega)
= i\rho_c^2 \omega G_{\alpha \beta} (q=0,
\omega+ i\epsilon)$.
Rotational invariance combined with fact that a magnetic field breaks
parity and time-reversal implies
\begin{eqnarray} \label{cond}
\sigma_{xx}& = \sigma_{yy}&= \rho_c^2{{i\omega [ -\rho_m\omega^2 + \Sigma
+
I (\omega)]}
\over { (\Sigma - \rho_m\omega^2 + I ( \omega))^2 -\rho_m^2 \omega^2
\omega_c^2}}
\nonumber \\
\sigma_{xy}&= -\sigma_{yx}& = \rho_c^2{{i\omega [ -i\rho_m\omega
\omega_c]}
\over { (\Sigma_1 - \rho_m\omega^2 + I ( \omega))^2 -\rho_m^2 \omega^2
\omega_c^2}}
\end{eqnarray}
In the absence of disorder one has $I=\Sigma=0$
in (\ref{cond}). $\sigma_{xx}$ vanishes in the
dc limit $\omega=0$ and has a $\delta-$function peak at
cyclotron frequency $\omega=\omega_c$. On the other hand
$\sigma_{xy}(\omega=0)={\rho_c
/B}$ and $\sigma_{xy}$ has a pole at $\omega=\omega_c$.
In the presence of disorder the crystal is pinned and
conductivities develop a new peak at the {\it pinning
frequency} $\omega=\omega_p$. Simultaneously,
there is an upward shift of the cyclotron resonance peak
from $\omega_c$ by a quantity of order $\omega_p$.

To obtain the full frequency dependence of
the conductivities one needs to compute
$I(\omega)$. This can be obtained
from the above equations which are valid
for all values of $B$. Here we present a solution
in the experimentally relevant
limit $\omega_c \gg {d / c}$. A typical plot of
$Re \sigma_{xx}$ obtained  by solving (\ref{iwn}) numerically  is shown in
Fig. 1.
Since $I(\omega=0)=0$  by definition in the pinned crystal,  the dc value
of
$\sigma_{xx}$ is still  zero but that of $\sigma_{xy}$
is zero in contrast to the pure case where
it was finite. The  peaks at the new  resonance frequencies
 have a finite height
and width due to disorder induced dissipation. The extent of
 this dissipation
is determined by  $I(\omega_n)$
 continued  to real frequencies.
Earlier results \cite{FL}  can be recovered
by setting $I(\omega_n)=0$ in all the equations. However, the presence of
the $I(\omega_n)$ term has
 many important physical consequences as will be discussed below.
Firstly,  in the absence of $I(\omega_n)$ the peaks would be delta
functions at $\omega^0_p$ and $\omega_c+\omega^0_p$
with $\omega^0_p=\Sigma/\omega_c$. In contrast,  here
the peaks  are centered
around  a frequency $\omega_p < \omega^0_p$  and this shift is given by
 $\Sigma$.
The peaks
 have a non-trivial structure
 and  are
asymmetric about the resonance frequencies as can be inferred from
(\ref{cond})
and  seen
in Fig. 1.
This invalidates the Lorentzian shape of the peaks
which was used to arbitrarily  broaden  the delta
functions in Ref.\onlinecite{FL}.
We note that
the peaks we obtain are  much narrower  than the Lorentizian broadened
ones.

For
frequencies $\omega \ll \omega_p$ and
$\omega_p\ll  \omega \ll \omega_c$,  analytical solutions
can be obtained. We find
\begin{eqnarray} \label {si1}
I(\omega_n) &=& {{\sqrt {2\rho_m\Sigma
 + {{\pi \rho_m^2\omega^2_c {\Sigma}^{1 \over 2}} \over {2 \sqrt{cd^2}}}}}}
\vert \omega_n \vert  ,\qquad \omega \ll \omega_p \nonumber \\
I(\omega_n) &=& {\Sigma \over 6} \log {{{\rho_m^2\omega_n^2 }\omega_c^2}
\over {d \Sigma^{ 3 \over 2}}} ,\qquad
\omega_p \ll \omega \ll \omega_c
\end{eqnarray}
Using (\ref{si1}) in
(\ref{cond}), we obtain the following low frequency behaviour
for $\omega \ll \omega_p$
\begin{eqnarray} \label{cxx}
Re\sigma_{xx}(\omega) &=
   &\rho_c^2{\sqrt {2\rho_m\Sigma
 + {{\pi \omega^2_c {\Sigma}^{1 \over 2}} \over {2 \sqrt{cd^2}}}}}
 {({\omega \over \Sigma})^2} \nonumber \\
Im\sigma_{xx}(\omega) &= \rho_c^2 &{\omega \over \Sigma} \nonumber \\
Re\sigma_{xy}(\omega)
 &=  &\rho_c^2 \rho_m\omega_c {({\omega \over \Sigma})^2} \nonumber \\
Im\sigma_{xy}(\omega) &\sim &\rho_c^2 \rho_m^{3 \over2} {{\omega_c
\omega^3} \over { \Sigma^{ 5 \over 2}}}
\end{eqnarray}
In the region 
$ \omega_p \ll \omega \ll \omega_c$
we find using  (\ref{si1})
\begin{equation} \label{cxx1}
Re\sigma_{xx}(\omega)  \sim {\rho_c^2 \over \rho_m^2} {\Sigma \over
{\omega_c^2 \omega}} \qquad
Re \sigma_{xy}(\omega)  \sim {\rho_c\over B}
\end{equation}
\noindent
Note that $Re\sigma_{xx}$ and $Re\sigma_{xy}$ are both
quadratic in  $\omega$.
Since the pinned WC has the characteristics proposed
for the HI, it seems unnecessary here to invoke the existence
of the HI as a new phase.
The results of  (\ref{cxx}) can be used to calculate the
 dielectric constant
$\epsilon(\omega) = {{\rm Im}\sigma_{xx}(\omega) / \omega}$.
Its dc value is given by
$\epsilon = \rho_c^2/\Sigma$. Thus the
dielectric constant is also a measure of the
characteristic frequency defined by disorder.

\begin{figure}
\centerline{\epsfig{file=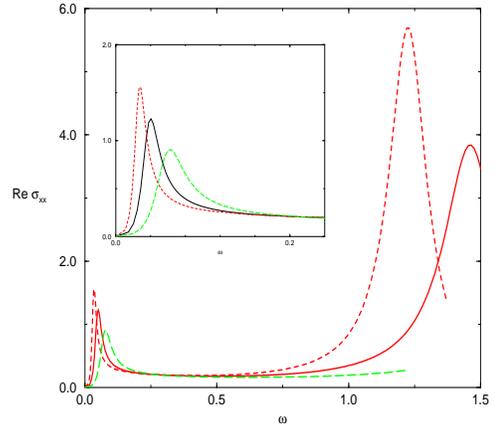,angle=180,width=7cm}}
\caption {$Re\sigma_{xx}(\omega)$ (in units of $b\rho_m /\rho_c$)
 as a
function of $\omega$ (in units of $\rho_c^3 / \rho_m^2 b$) 
where
$b= cAa^{2}\Delta\rho_m^{-2} ( e / \hbar)^3$ 
for the case $r_f < l_c$ for different values of the field $B$ meausred in
units of $\rho_c^2 / \rho_m b$.
The dashed line represents $B=0.33 $,
the full line  $B=0.4$ and the long dashed line
$B=0.5$. The inset is a magnification of the pinning peak.}
\label{fig1}
\end{figure}

Calculating the   resistivities   given by
$\rho_{\alpha \beta}= -\sigma_{\alpha \beta}/( \sigma_{xx}^2 +
\sigma_{xy}^2 )$
we find that the pinned crystal is  indeed  insulating i.e.,
$\rho_{xx}(\omega=0)= \infty $.
More importantly, the  Hall resisitivity
$\rho_{xy}$ turns out to be  independent of $\omega$  and $T$
and has the same value as that in the pure system
$\rho_{xy}(\omega)= {B / \rho_c}$.
A similar  result was argued to hold
at $T=0$ in Ref.\onlinecite{NLM}.
At $T>0$ it would be necessary to go beyond the
GVM approximation to ascertain whether
$\rho_{xy}$ still sticks to its classical
value. Indeed the GVM misses soliton like
excitations which are known to be important for
finite $T$ physics \cite{TP,LL}.

It is interesting to calculate the  field
dependences of  the above quantities. This  is of
direct  experimental relevance and we use our results to calculate the
same.
The field dependence of the pinning peak (whose width is naively of
$O(\omega_p)$) is governed by $\Sigma$
whose value  is in turn dictated by the relative sizes of the length
scales
$r_f$ and $l_c$. Traditionally, in the context of CDW,
$\Sigma$ has been related to the Fukuyama-Lee length $R_a$
at which relative displacements are of order $a$,
as $\Sigma \sim R_a^{-2}$. However for the present case
such a connection does not hold, because disorder can
a priori vary at scales much smaller than $a$ unlike in CDW.
This is similar to the situation in vortex systems
where pinning is controlled by the Larkin length $R_c$,
defined as the scale
below which the physics of (\ref{decadix}) can be described perturbatively
by a model
where uncorrelated gaussian random forces of strength
$\Delta_f = \sum_{\bf K} {\bf K}^2 \Delta_K$ act independently on each electron.
Within this model $R_c$ is given
by $\langle [{\bf u}({\bf R}_c) -{\bf u}({\bf 0})]^2\rangle_{f}
= \rm{max} [r_f ^2, l_c^2 ] \equiv \xi^2_0$. When
$R_c \gg a$ the crystal is pinned collectively.
In this regime  using the GVM  one finds
$\Sigma = c A R_a^{-2} (a/\xi_0)^6$. Here
$A$ is an overall constant, $R_a \sim {{\rho_m a^2} / {\sqrt \Delta}}$ and
$\xi_0= max[r_f,l_c]$. This corresponds to $\Sigma  \sim c R_c^{-2}$,
implying that $l \sim R_c$ (defined above),
which shows that the length scale determining
the peak in the  conductivity is  $R_c$ and not $R_a$.
It is important to distinguish  between these two lengths
since $R_c$ can have an explicit dependence on the
magnetic field. This yields two very different regimes.
One is $r_f < l_c$,
 which gives $\Sigma= b B^3$ ($b= cAa^{2}\Delta\rho_m^{-2} ( e / \hbar)^3$
) hence
$\omega_p(B) \propto B^2$ and the pinning peak moves up and
broadens with increasing field. This is the case illustrated in  Fig. 1
where
 $Re\sigma_{xx}$ has been plotted
 for various values of  $B$. The second regime is
$r_f >l_c$ leading to  a $\Sigma$ independent of $B$ and
$\omega_p(B) \propto  B^{-1}$ . Thus
the pinning peak moves towards the origin and gets narrower
with increasing field. For CDW in a magnetic field $r_f \sim a$
and one is always in the second regime. In these two regimes
the height of the pinning peak
{\it decreases}  as $B^{-1}$  with increasing field.
For the case $r_f \ll l_c$,  $\omega_p$
does not increase indefinitively with $B$, rather there is a
crossover to another regime when $R_c \sim a$ where
single particle pinning effects are dominant. Here the correspondence
between $R_c$ and $l$ and hence $\Sigma$ no longer holds. One finds then
that $\Sigma \propto B^{3 \over 2}$ and
$\omega_p \propto B^{ 1\over 2}$ provided $l_c \gg r_f$.
In contrast the  peak at
$\omega_c + \omega_p$  always moves upwards with increasing $B$.
A summary of the  results in various  regimes  is given
in Table~\ref{table1}.

Another important measurable quantity is
the threshold electric field $E_T$ necessary for the
crystal to slide. This again  shows the interplay
between $l_c$ and $r_f$. Using collective
pinning arguments \cite{fn} one gets
$E_T= {c } R_c^{-2} \xi_0$ in the regime $R_c \gg a$.
The threshold field has the same regimes as above and
the  field dependence are shown  in
Table~\ref{table1}. Note that for $r_f > l_c$
$E_T$ is independent of the field (as for CDW).
Since both $E_T$ and $\omega_p$ are related to $R_c$
one has $E_T \propto \omega_p$ but with a prefactor
depending on $\xi_0$ and not on $a$ as given by previous
CDW estimates.
When $R_c \leq a$, we enter the regime of single particle pinning.
Using $\Sigma \propto B^{3 \over 2}$, the threshold field is now
$E_T \propto  \Sigma l_c \propto B$ \cite{footnote2}.
Finally, due to the variation of $\Sigma$ with the field
the dielectric constant $\epsilon$ will
exhibit  the behaviours shown in Table~\ref{table1}.
Therefore in addition to detailed frequency measurements of
the conductivity, measurements of
the $B$ dependences of the dielectric constant could serve
as an experimental signature for the WC.

Some of the existing experimental results can be interpreted within
our theory. Contrary  to previous estimates of the pinning
frequency, it allows   a scenario where the   pinning
frequency  {\it increases} with the field as was seen in recent
experiments \cite{CC}.  A simultaneous  increase in
$E_T$ vs. $B$ is observed  which is in qualitative agreement with
the above predictions. Our theory also predicts that the
Hall resistance takes its classical value which is observed
experimentally \cite{SL}.
However,  many problems remain both theoretically and in
comparison with experiments. Experimentally the
peak height in $\sigma(\omega)$ seems to increase
with $B$ which we cannot account for at present.
Some experiments \cite{SL} seem to report
a different behaviour for the conductivity
(see however \cite{footnote1}). Finite temperature
effects also  need to be understood since many
experiments are performed in the d.c. limit at
$T>0$. Strong disorder effects have also
to be understood. Both problems require a
careful treatment of the topological defects
and solitons which are  beyond the scope of the
present study. While a phase transition similar to the
one occuring in 3d vortex lattices
\cite{TGPL} is unlikely in
$d=2$, one expects a marked crossover between a
weakly pinned WC and a strongly pinned one.

To conclude, we have developed a comprehensive theory for the  WC pinned
by weak disorder. In addition to
detailed frequency dependences of the real and imaginary
parts of the conductivites, we have obtained the magnetic field
dependences of various dynamical quantities.
We find that the magnetic field not only confines the
electrons but also plays a crucial role in determining
the response of the system
to disorder. This dynamical effect, not captured by
previous static approximations, allows the
possibility of observing novel field dependences.

We thank  F.I.B. Williams for enlightening remarks.

\narrowtext
\begin{table}
\begin{tabular}{lcccc}
Regime  &  $\Sigma$ &  $\omega_p$ &  $E_T$ & $\epsilon$ \\
\tableline
$r_f > l_c$  &  $B^0$ &  $B^{-1}$ &  $B^0$ &  $B^0$ \\
$r_f < l_c$  &  $B^3$ &  $B^{ 2}$  &  $B^{5 \over 2}$ &  $B^{-3}$\\
$R_c < a$  &  $B^{3 \over2}$  &  $B^{ 1 \over 2}$ &  $B$
& $B^{-{3 \over 2}}$ \\
\tableline
\end{tabular}
\caption{Magnetic field dependences of various dynamical quantities.
\label{table1}}
\end{table}
\end{multicols}

\begin{thebibliography}{99}
\bibitem[*]{cnrs} Laboratoire associ\'e au CNRS.
\bibitem{EW} E. Wigner, Phys. Rev. {\bf 46}, 1002(1934).
\bibitem{WC1} E.Y. Andrei, {\it {et al}}, Phys. Rev. Lett. {\bf
60},2765(1988).

\bibitem{WS} R.L Willet {\it et al}, Phys. Rev. Lett. {\bf 65},633 (1990).
\bibitem{NC1} F.I.B. Williams, {\it et al}, Phys. Rev. Lett. {\bf 66},
3285
(1991);
 F. Perruchot thesis, Ecole Polytechnique, Paris, 1995.

\bibitem{WC2} V.J. Goldman {\it et al},
Phys. Rev. Lett. {\bf 65}, 2189 (1990).

\bibitem{BWC} H.C. Manoharan, {\it et al},
Phys. Rev. Lett. {\bf 77}, 1813 (1996).

\bibitem{SL}V.J. Goldman, M. Shayegan and D. Tsui, Phys. Rev. Lett.
{\bf 61},881 (1988); T. Sajoto {\it et al},
{\it ibid} {\bf 70}, 2321 (1993).


\bibitem{KT} I.V. Kukushkin and V.B. Timofeev, Phys- Uspekhi {\bf 36}, 549
(1993).
\bibitem{Klit} A.A. Shashkin {\it et al}, Phys. Rev. Lett. {\bf 73},
3141 (1994).

\bibitem{KLZ} S.C. Zhang, S. Kivelson and D.H. Lee, Phys. Rev. Lett.
{\bf 69}, 1252 (1992).

\bibitem{TGPL} T. Giamarchi and P. Le Doussal
Phys. Rev. B {\bf  52}, 1242 (1995).

\bibitem{FL}  H. Fukuyama and P. Lee, Phys. Rev. {\bf B} 18, 6245 (1978).

\bibitem{NLM} B.G.A. Normand, P.B. Littlewood and A.J. Millis,
Phys. Rev. B {\bf 46}, 3920 (1992);
 X. Zhue, P.B. Littlewood and A.J. Millis,
{\it ibid} {\bf 50}, 4600 (1994).

\bibitem{WKS} U. Wulf, J. Kucera and E. Sigmund Phys. Rev. Lett. {\bf 77}
2993 (1996).

\bibitem{SL1} Y.P. Li {\it et al},
 Solid State Comm. {\bf 95}, 619 (1995).

\bibitem {CC} C.C. Li {\it et al}, preprint; F.I.B. Williams,
private communication.

\bibitem{BM} L. Bonsall and A.A. Maradudin, Phys. Rev. B {\bf 15}, 1959
(1977).

\bibitem{TP} T. Giamarchi and P. Le Doussal
Phys. Rev. B {\bf 53} 15206 (1996).

\bibitem{future} R. Chitra, T. Giamarchi and P. Le Doussal, manuscript
in preparation.

\bibitem{LA} A.I. Larkin, Zh. Eksp. Teor. Fiz. {\bf 58},1466 (1970).

\bibitem{fn} A.I. Larkin and Y.M. Ovchinnikov, J. Low Temp. Phys. {\bf
34},
409 (1979).

\bibitem{NV} D.R. Nelson and V.M. Vinokur, Phys. Rev. B {\bf 48},
13060(1993).

\bibitem{LL} A. I. Larkin and P. A. Lee Phys. Rev. B {\bf 17} 1596
(1978).

\bibitem{footnote1} They
have reported $Re \sigma_{xx}$ and $Im\sigma_{xx}  \propto \omega$.
However, this linear behaviour in  both  $Re\sigma_{xx}$ and  
$Im\sigma_{xx}$
violates analyticity properties and more work is needed to clarify
this point before a comparison with theory can be made.

\bibitem{footnote2}
This holds because here the ``single particule localization length''
(see \cite{TGPL,NV}) $l_{\perp} \sim l_c$. There could be an  additional
regime where $r_f \ll l_{\perp} \ll l_c$ where arguments similar
to \cite{NV} can be made.

\end{thebibliography}
\end{document}